\documentclass[10pt]{article}
\usepackage{epsf}
\usepackage[dvips]{graphicx}
\usepackage{amssymb}

\begin{document}

\begin{center}
{\bfseries NUCLEUS-NUCLEUS COLLISIONS AT LOW ENERGIES. THE EFFECTS
FROM NON VACUUM EXCHANGE}

\vskip 5mm

N.V. Radchenko$^{\dag}$, A.V. Dmitriev$^{\ddag}$

\vskip 5mm

{\small {\it Novgorod State University }
\\
$\dag$ {\it E-mail: Natalya.Radchenko@novsu.ru }
\\
$\ddag$ {\it E-mail: alexandrvdmitriev@gmail.com }}
\end{center}

\vskip 5mm

\begin{abstract}
Experimental data on total and differential elastic cross sections
for $p+p(\bar{p})$, $n+p(\bar{p})$, $K^\pm+p$, $K^\pm+n$,
$\pi^\pm+p$ starting from energy 3.5 GeV in CMS are used to
determine parameters of vacuum contribution and parameters of
basic non vacuum reggeons: $f$, $\omega$, $\rho$ and $A_2$. It is
argued that non vacuum contributions to proton-proton and
proton-neutron collisions correspond to spectrum in which baryon
number is moved from the fragmentation region to central region in
rapidity space. In this case it is possible that chemical
potential is increased in central region of spectrum of
nucleus-nucleus interaction at low energies. This effect might be
important for facilities FAIR and NICA.
\end{abstract}

\vskip 10mm

\section{Introduction}
Processes with multiple production at high energies are
intensively studied both experimentally and theoretically. At that
time behavior of multiple processes at low energies
$\sqrt{s}\approx 3.5\div10$~GeV is known much less. Multiple
processes at low energies are dominated by contributions of non
vacuum reggeons. While ``pomeron physics'' is considerably well
explored basing on QCD (\cite{bib1} -- \cite{bib6}, also
see~\cite{bib7} and references therein) clear QCD based picture of
multiple processes, associated with non vacuum reggeons is still
missing.

At low energies ($\sqrt{s}=3.5\div10$~GeV in center-of-mass
system) non vacuum exchange can give up to half of total cross
sections value in $pp$ and $pn$ interactions. So its analysis is
important for better understanding of nucleus-nucleus scattering
processes, which will occur at facilities FAIR and NICA. Unlike
works~\cite{bib8}, \cite{bib9} we consider four  basic non vacuum
reggeons. The extensive data set of the Particle Data
Group~\cite{bib10} at low energies gives possibility to do it.
Firstly this approach is more correct. Secondly knowledge of
quantum non vacuum exchange numbers can be essential for
signatures of different processes which take place when passing of
nucleus through nucleus.

In present work we will study out which ``color diagrams''
correspond to non vacuum reggeons. Analysis of these color
diagrams will show that at low energies baryon number increase in
central region of produced particles spectrum. For nucleus-nucleus
collisions it means that baryon chemical potential is increased in
central region of spectrum. This may lead to discovery of phase
transition to quark-gluon plasma at energies of facilities NICA
and FAIR.

\section{Total cross sections of various reactions}
In the Low Constituents Number Model (LCNM)~\cite{bib5},
\cite{bib6} total cross sections can be presented as sum of two
parts -- vacuum contribution rising as logarithm of full energy
squared and non vacuum contribution. Vacuum contribution contains
constant part and terms proportional to logarithm of full energy
and to logarithm of full energy squared. These terms correspond to
contributions of one and two additional gluons in initial state.

We consider exchange of four basic non vacuum reggeons: $f$-,
$\rho$, $\omega$ and $A_2$-mesons. Therefore formulas for total
cross sections can be written as follows.
\begin{equation}\label{t1}\begin{array}{lcl}
\sigma_{tot}^{pp(\bar{p})}&=&\sigma_0^{pp}+\sigma_1^{pp}\ln
s+\sigma_2^{pp}\ln^2 s+\\&&g_f^p g_f^p s^{-\Delta_f}\pm g_\rho^p
g_\rho^p s^{-\Delta_R}\pm g_\omega^p g_\omega^p
s^{-\Delta_R}+g_{A_2}^p g_{A_2}^p s^{-\Delta_R}
\\[3mm]
\sigma_{tot}^{np(\bar{p})}&=&\sigma_0^{pp}+\sigma_1^{pp}\ln
s+\sigma_2^{pp}\ln^2 s+\\&&g_f^p g_f^p s^{-\Delta_f}\pm g_\rho^p
g_\rho^p s^{-\Delta_R}\mp g_\omega^p g_\omega^p s^{-\Delta_R}-
g_{A_2}^p g_{A_2}^p s^{-\Delta_R}
\\[3mm]
\sigma_{tot}^{K^\pm p}&=&\sigma_0^{Kp}(1+\delta_1^{pp}\ln
s+\delta_2^{Kp}\ln^2 s)+\\&&g_f^K g_f^p s^{-\Delta_f}\mp g_\rho^K
g_\rho^p s^{-\Delta_R}\mp g_\omega^K g_\omega^p s^{-\Delta_R}+
g_{A_2}^K g_{A_2}^p s^{-\Delta_R}
\\[3mm]
\sigma_{tot}^{K^\pm n}&=&\sigma_0^{Kp}(1+\delta_1^{pp}\ln
s+\delta_2^{Kp}\ln^2 s)+\\&&g_f^K g_f^p s^{-\Delta_f}\pm g_\rho^K
g_\rho^p s^{-\Delta_R}\mp g_\omega^K g_\omega^p s^{-\Delta_R}-
g_{A_2}^K g_{A_2}^p s^{-\Delta_R}
\\[3mm]
\sigma_{tot}^{\pi^\pm p}&=&\sigma_0^{\pi p}(1+\delta_1^{pp}\ln
s+\delta_2^{\pi p}\ln^2 s)+\\&&g_f^\pi g_f^p s^{-\Delta_f}\mp
g_\rho^\pi g_\rho^p s^{-\Delta_R}
\end{array}
\end{equation}

The numerical values of all parameters are presented in Tables~1
and 2. The examples of fitting are shown for $np(\bar{p})$ in
Figure~1. Experimental data were taken from the Particle Data
Group~\cite{bib10}.

\begin{figure}[!h]
\centerline{
\includegraphics[scale=0.6]{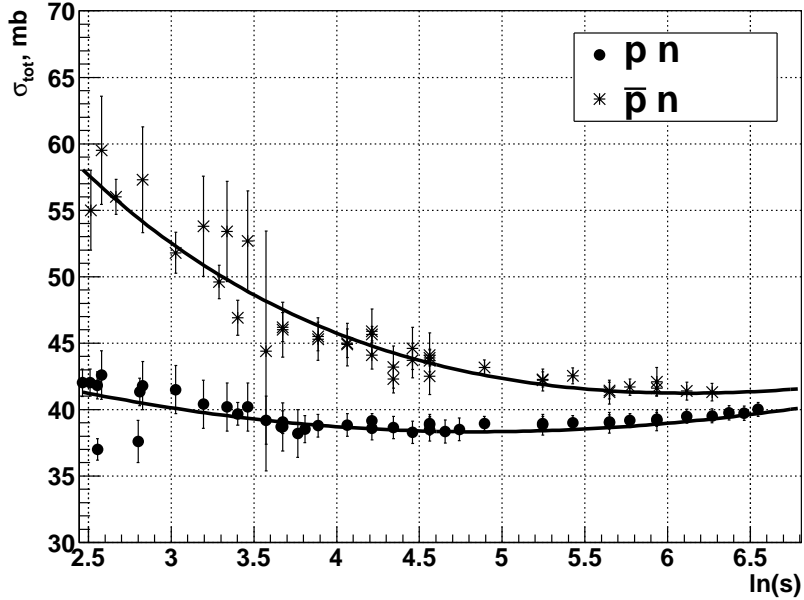}}
\caption{Fitting of total cross sections}
\end{figure}


\begin{table}[!h]
\caption{Values of vacuum parameters}
\begin{tabular}{|c|c|c|c|c|}
\hline $\sigma_0$, mb&$\sigma_1$, mb&$\sigma_2$,
mb&$\delta_1=\sigma_1/\sigma_0$&$\delta_2=\sigma_2/\sigma_0$\\\hline
\multicolumn{5}{|c|}{$pp$, $p\bar{p}$, $np$, $n\bar{p}$}\\\hline
$21.63\pm1.94$&$0.84\pm0.39$&$0.18\pm0.02$&$0.04\pm0.02$&$0.008\pm0.001$\\\hline
\multicolumn{5}{|c|}{$K^\pm p$, $K^\pm n$}\\\hline
$10.43\pm0.88$&$0.40\pm0.19$&$0.15\pm0.02$&$0.04\pm0.02$&$0.014\pm0.002$\\\hline
\multicolumn{5}{|c|}{$\pi^\pm p$}\\\hline
$11.52\pm0.93$&$0.45\pm0.21$&$0.17\pm0.02$&$0.04\pm0.02$&$0.015\pm0.001$\\\hline
\end{tabular}
\end{table}

\begin{table}[!h]
\caption{Values of non vacuum parameters}
\begin{tabular}{|c|c|c|c|}
\hline vertex& $g^p$, mb${}^{1/2}$ ($p$, $n$, $\bar{p})$&$g^K$,
mb${}^{1/2}$ ($K^\pm$)&$g^\pi$, mb${}^{1/2}$ ($\pi^\pm$)\\\hline
$f$-meson&$7.85\pm0.20$&$2.46\pm0.15$&$4.28\pm0.11$\\\hline
$\rho$-meson&$1.51\pm0.21$&$2.05\pm0.35$&$3.75\pm0.35$\\\hline
$\omega$-meson&$6.00\pm0.14$&$1.88\pm0.06$&\\\hline
$A_2$-meson&$1.33\pm0.24$&$1.64\pm0.38$&\\\hline
\end{tabular}
\end{table}

\section{Dual topological diagrams}
Hadrons interactions with exchange of non vacuum reggeon
correspond to soft processes with flavor transfer in $t$ channel.
Since only slow partons softly interact with each other, initial
state configurations where one of valence quarks has low momentum
are very essential.

In frame of dual resonance model (\cite{bib11} -- \cite{bib14})
slowing of quark and reggeon exchange is depicted as dual diagram
in Fig.~2a (we consider $\pi^+\pi^-$ scattering).

\begin{figure}[!h]
\centerline{
\includegraphics[scale=0.6]{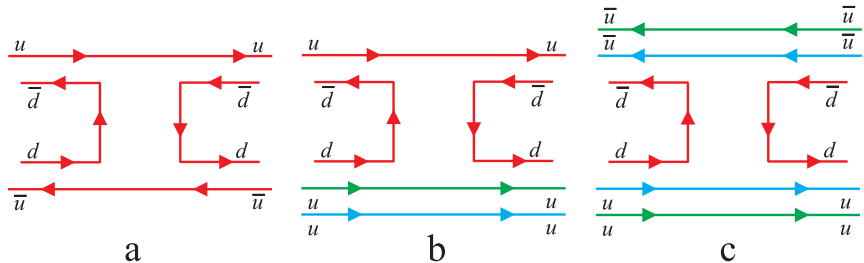}}
\caption{a) Dual diagram for $\pi^+\pi^-$ scattering. b) Dual
diagrams for $\pi^+p$ interaction. Completely analogous diagram
can be drawn for $\pi^-p$ scattering. c) Dual diagram for
$p\bar{p}$ scattering.}
\end{figure}

In this approach hadrons (mesons) represent string with quark and
antiquark at its endpoints.  When moving in 4-dimensional
space-time string sweeps out 2-dimensional surface. Diagram in
Fig.~2a shows elastic interaction. String endpoints of initial
state merge and further one quark string moves in $s$ channel,
which then splits into two strings. Consequently elastic
scattering amplitude appears, which constitutes smooth
2-dimensional surface. The same 2-dimensional surfaces correspond
to amplitudes of $n$ particles production. More complicated
structure, in which hollow cylinder is glued to poles that are
swept out by strings of initial hadrons, corresponds to pomeron.

Dual resonance model gives independent inference of reggeon
diagram technique, which does not use expansion in colorless
particles of amplitudes or parton wave functions. The AGK
theorem~\cite{bib15} may be obtained also in frame of this
approach~\cite{bib7}. Dual diagrams for $\pi^+p$ and $p\bar{p}$
interactions are given in Fig.~2b and 2c.

In case of $\pi^{\pm}p$ interaction there is stage when only quark
string moves  in $s$ channel, similarly to diagram in Fig.~2a.
Endpoints of this string are quark and diquark.

In case of $p\bar{p}$ interaction region is swept out by string
with quark and diquark at its endpoints.

In what follows we will construct color diagrams for non vacuum
reggeons using two results, obtained from consideration of dual
diagrams DRM.

1. In $s$ channel of color diagrams there must exist quark string,
which is not divided into several parts. This string has quarks
(antiquarks) and diquarks (antidiquarks) at its endpoints.

2. In $t$ channel elastic amplitudes, describing reggeon
contributions, must have quark-antiquark pair.

\section{Color diagrams for nucleon-nucleon scattering}
There are no dual diagrams for nucleon-nucleon collisions (we will
consider proton-proton scattering as example). This process is
described by so-called twist diagrams, in which scattering of
slowed quarks takes place, but not annihilation of quark and
antiquark. In the first approximation of DRM such diagrams do not
contribute to imaginary part of nucleon-nucleon scattering. So non
vacuum Regge contributions which are present in meson-nucleon and
antinucleon-nucleon interactions must not exist in case of
nucleon-nucleon interactions. But these contributions are visible
in experimental data and parameters of their Regge trajectories
(intercepts and slopes) coincide with  Regge trajectories
parameters of $\pi^\pm p$, $K^\pm p$, $p\bar{p}$, $n\bar{p}$
collisions.

Therefore there must exist a diagram which fulfills requirements
of dual diagrams. We have found two diagrams of process with one
quark string in $s$ channel, Fig.~3.

\begin{figure}[!h]
\centerline{
\includegraphics[scale=0.5]{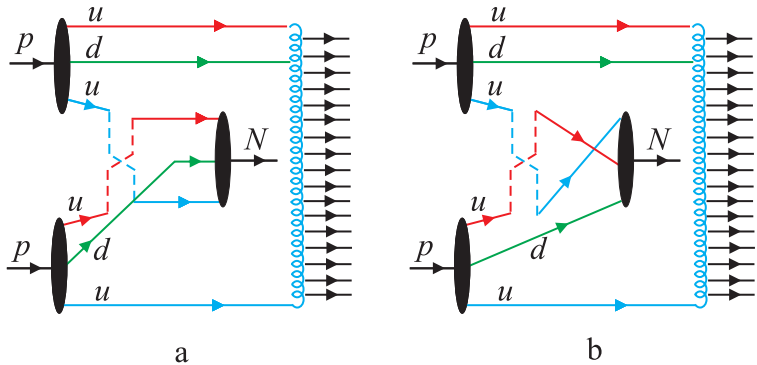}}
\caption{$pp$ interaction.}
\end{figure}

In diagram from Fig.~3a quarks move forward after scattering, in
diagram from Fig.~3b they move backward. In order to form one
quark string in $s$ channel one of protons must be taken in
configuration with slowed diquark. String in $s$ channel has quark
and diquark at its endpoints. Since this string breaks out into
secondary hadrons, then slow state with three quarks also forms
colorless state -- some baryonic resonance.

\section{Conclusions}
Thus we can argue that leading non vacuum reggeons in
nucleon-nucleon scattering (proton-proton, proton-neutron and
neutron-neutron) result in translocating of baryons from
fragmentation region to central region of secondary hadrons
spectrum.

The obtained result means that in nucleus-nucleus scattering at
low and intermediate energies baryon number may increase in
central region of secondary hadrons spectrum. This may help in
discovering quark-gluon plasma effects in facilities NICA and
FAIR.

We have come to conclusions by studying structure of color
diagrams for non vacuum reggeons. Evidently, further detailed
analysis is necessary. Though many important results were obtained
only from structure of diagrams in $\lambda\varphi^3$ theory. In
particular, the first proof of the AGK theorem was derived exactly
from analysis of ladder diagram structure in $\lambda\varphi^3$.

One of authors (N.V. Radchenko) gratefully acknowledges financial
support by grant of Ministry of education and science of the
Russian Federation, federal target program ``Scientific and
scientific-pedagogical personnel of innovative Russia'', grant
P1200.


\begin{thebibliography}{99}
\bibitem{bib1}
E.A. Kuraev, L.N. Lipatov and V.S. Fadin, Sov. Phys. JETP
\textbf{44} (1976) 443.

\bibitem{bib2}
I.I. Balitsky, L.N. Lipatov, Sov. J. Nucl. Phys. \textbf{28}
(1978) 822.

\bibitem{bib3}
V.A. Abramovsky, O.V. Kancheli, Pisma Zh. Eksp. Teor. Fiz.
\textbf{31} (1980) 566.

\bibitem{bib4}
V.A. Abramovsky, O.V. Kancheli, Pisma Zh. Eksp. Teor. Fiz.
\textbf{32} (1980) 498.

\bibitem{bib5}
V.A. Abramovsky, N.V. Radchenko,   Particles and Nuclei, Letters
\textbf{6} (2009) 607.

\bibitem{bib6}
V.A. Abramovsky, N.V. Radchenko,  Particles and Nuclei, Letters
\textbf{6} (2009) 717.

\bibitem{bib7}
V.A. Abramovsky, E.V. Gedalin, E.G. Gurvich, O.V. Kancheli,
Inelastic interaction at high energies and chromodynamics,
Tbilisi, Metzniereba, 1986 (in russian).

\bibitem{bib8}
A. Donnachie, P.V. Landshoff, Phys. Lett. \textbf{B 296} (1992)
227.

\bibitem{bib9}
J.R. Cudell  et al., Phys. Rev. \textbf{D 61} (2000) 034019.

\bibitem{bib10}
C. Amsler  et al., Review of Particle Physics, Phys. Lett.
\textbf{B 667} (2008) 1.


\bibitem{bib11}
C. Rebbi, Phys. Rept. \textbf{C12} (1974) 2.

\bibitem{bib12}
S. Mandelstam, Phys. Rept. \textbf{C13} (1974) 259.

\bibitem{bib13}
J. Scherk, Rev. Mod. Phys. \textbf{47} (1975) 123.

\bibitem{bib14}
G.F. Chew,  C. Rosenzweig, Phys. Rept. \textbf{41} (1978) 263.

\bibitem{bib15}
V.A. Abramovsky, V.N. Gribov, O.V. Kancheli, Sov. J. Nucl. Phys.
\textbf{18} (1974) 308.



\end{thebibliography}
\end{document}